\documentclass[sort&compress,5p]{elsarticle}
\usepackage{graphicx,natbib,amssymb}
\usepackage[pdftex,colorlinks,citecolor=blue,bookmarks]{hyperref}
\usepackage{bm}
\usepackage{amsmath}

\begin{document}
\title{Symmetry conserving configuration mixing method with cranked states}

\author[UAM]{Marta Borrajo}
\ead{marta.borrajo@uam.es}
\author[UAM]{Tom\'as R. Rodr\'iguez} 
\ead{tomas.rodriguez@uam.es}
\author[UAM]{J. Luis Egido} 
\ead{j.luis.egido@uam.es}
\address[UAM]{Departamento de F\'isica Te\'orica, Universidad
  Aut\'onoma de Madrid, E-28049 Madrid, Spain}
\begin{abstract}
We present the first calculations of a symmetry conserving configuration mixing method (SCCM) using time-reversal symmetry breaking Hartree-Fock-Bogoliubov (HFB) states with the Gogny D1S interaction. The method includes particle number and tridimensional angular momentum symmetry restorations as well as configuration mixing within the generator coordinate method (GCM) framework. The nucleus $^{32}$Mg is chosen to show the performance and reliability of the calculations. Additionally, $0^{+}_{1}$, $2^{+}_{1}$ and $4^{+}_{1}$ states are computed for the magnesium isotopic chain, where a noticeable compression of the spectrum is obtained by including cranked states, leading to a very good agreement with the known experimental data. 
\end{abstract}
\begin{keyword}
Beyond-mean-field theories, GCM, Time-reversal symmetry breaking, Density functionals
\PACS{21.60.Jz,21.10.Re,27.30.+t}
\end{keyword}
\maketitle
\begin{figure}[t]
\begin{center}
  \includegraphics[width=\columnwidth]{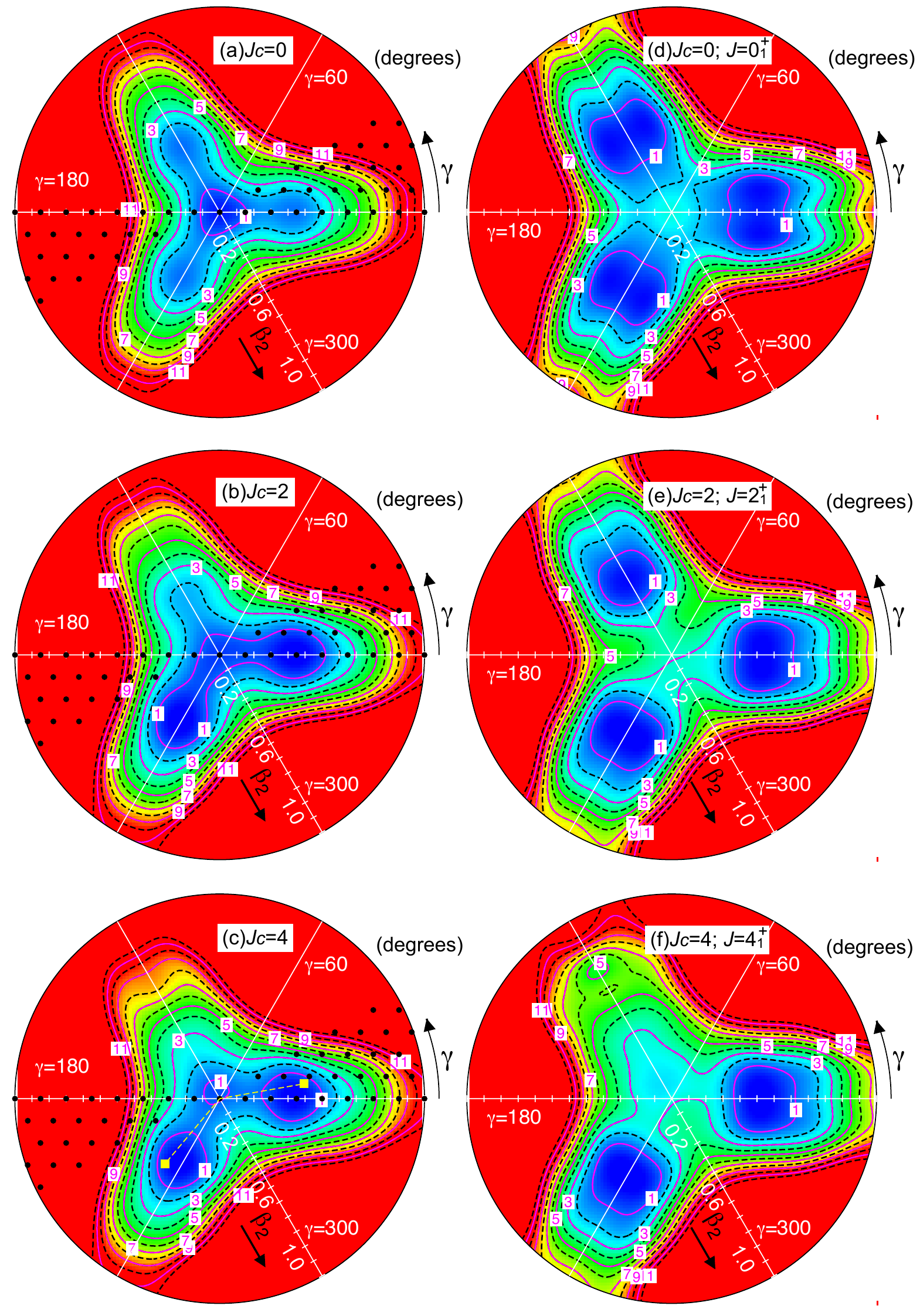}
\end{center}
\caption{(color online) (a)-(c) PN-VAP and (d)-(f) Particle number and angular momentum projected potential energy surfaces for different values of the cranking angular momentum $J_{c}$ for the nucleus $^{32}$Mg. Gogny D1S interaction is used here. The contour plots are separated in energy by 1.0 MeV. Each PES is normalized to the energy of their corresponding minima, i.e., (a) -249.902 MeV, (b) -247.910 MeV,  (c) -246.789 MeV, (d) -252.924 MeV, (e) -252.021 MeV and (f) -250.463 MeV. The black bullets in (a)-(c) are the states included in the GCM calculation while the yellow squares in (c) are the states analyzed in detail in Figs.~\ref{Figure3}-~\ref{Figure4}.}
  \label{Figure2}
\end{figure}
A trustworthy description of the spectra of the atomic nuclei is one of the main goals of low-energy nuclear structure theory. The interacting shell model~\cite{RMP_77_427_2005,PPNP_47_319_2001} is likely the most widely used and successful tool to compute accurately low-lying spectroscopic properties. However, shell model applications are limited to regions not far away from shell closures where manageable valence spaces can be defined. 
On the other hand, microscopic self-consistent mean field methods (SCMF)~\cite{RMP_75_121_2003} based on nuclear energy density functionals such as Skyrme, Gogny and/or Relativistic Mean Field (RMF) can be in principle used throughout the whole nuclear chart. In order to apply these methods to the study of nuclear spectra, they have to be extended by including beyond-mean-field (BMF) correlations. In particular, symmetry conserving configuration mixing methods (SCCM) are the most natural extensions of SCMF approaches and have shown a  fair performance in describing qualitatively not only nuclear spectra, but also ground state properties, electromagnetic transitions and decays.
Unfortunately, quantitative accurate predictions have not been reached so far, mainly due to the lack of time-reversal symmetry breaking intrinsic states within the existing implementations of the SCCM methods. 
In this letter we present an extension of the SCCM framework, based on the generator coordinate method (GCM) with particle number and triaxial angular momentum projections, that includes cranking intrinsic states. In the numerical applications we use the finite range density dependent Gogny interaction (D1S parametrization~\cite{Berger84}). 
In the earliest implementations of the SCCM method only axially symmetric intrinsic states were considered~\cite{NPA_671_145_2000,NPA_709_201_2002,PRL_99_062501_2007,PRC_74_064309_2006,PLB_704_520_2011}. 
A major breakthrough towards a better description of the nuclear spectra within the SCCM framework was the inclusion of the triaxial degree of freedom~\cite{PRC_78_024309_2008,PRC_81_064323_2010,PRC_81_044311_2010}. 
A further step forward was the first implementation of a SCCM method based on a Skyrme pseudo-potential for odd nuclei by  B. Bally  who obtained very promising results in describing the benchmark nucleus $^{25}$Mg~\cite{PRL_113_162501_2014}.

On the other hand, the angular momentum projection with the energy density functionals mentioned above is performed after the energy variation. Therefore, the consideration of only intrinsic wave functions with zero angular momentum content $ (<J_x>=<J_y>=<J_z>=0)$
in the  the current SCCM calculations tends to favor the ground states with respect to other excited states and a stretching in the spectra is usually found with respect to the experimental values.
The addition of time-reversal symmetry breaking intrinsic states $ (<J_x>\ne 0)$ obtained by the cranking procedure will thus increase the variational space for excited states and will provide a better description of the spectrum. Pioneering angular momentum projection of cranking states have been reported with schematic pairing plus quadrupole interactions~\cite{NPA_385_14_1982,NPA_435_477_1985,PRC_59_135_1999} and with Skyrme energy density functionals~\cite{PRC_29_1056_1984,PRC_76_044304_2007}. However, neither configuration -shape- mixing nor, in the case of Skyme interactions, pairing correlations, were taken into account. The aim of this letter is to present the first results of the extension of the SCCM method described in detail in Ref.~\cite{PRC_81_064323_2010} (and references therein), including now time-reversal-symmetry breaking intrinsic states introduced through cranking calculations.

The starting point is the construction of a set of intrinsic many-body states having different deformations and intrinsic angular momentum. Such states, $|\beta_{2},\gamma,J_{c}\rangle\equiv|\rangle$, have the structure of Hartree-Fock-Bogoliubov (HFB) states~\cite{RingSchuck} and are found by minimizing the particle number projected HFB energy\footnote{For the sake of simplicity, we will write down throughout the text any energy kernel as an expectation value of a hamiltonian operator. However, Gogny interactions contain a density-dependent term which prevents such a notation rigorously~\cite{PRC_79_044318_2009}. Nevertheless, this term is handled separately in a similar fashion as in Refs.~\cite{NPA_709_201_2002,PRC_81_064323_2010}, and the following notation can be still used.}, i.e.:
\begin{equation}
E'_{J_{c}}(\beta_{2},\gamma)=\frac{\langle\hat{H}P^{N}P^{Z}\rangle}{\langle P^{N}P^{Z}\rangle}-\omega_{J_{c}}\langle\hat{J}_{x}\rangle-\lambda_{q_{20}}\langle\hat{Q}_{20}\rangle-\lambda_{q_{22}}\langle\hat{Q}_{22}\rangle
\label{vap_equat}
\end{equation}
where $P^{N(Z)}$ is the neutron (proton) number projection operator~\cite{RingSchuck}. This is the so-called  variation after particle number projection method (PN-VAP)~\cite{NPA_696_467_2001}.
The first term in the r.h.s. of Eq. \ref{vap_equat} is the particle number projected energy and the last terms correspond to the constraints on the cranking angular momentum $J_{c}$ and on the quadrupole deformation of the system $(\beta_{2},\gamma)$. The Lagrange multipliers $\omega_{J_{c}}$, $\lambda_{q_{20}}$ and $\lambda_{q_{22}}$ ensure the conditions:
\begin{equation}
\langle\hat{J}_{x}\rangle=\sqrt{J_{c}(J_{c}+1)};\,\langle\hat{Q}_{20}\rangle=q_{20};\langle\hat{Q}_{22}\rangle=q_{22}
\end{equation}
where $\hat{J}_{x}$ is the $x$-component of the angular momentum operator and $\hat{Q}_{2\mu}$ with $\mu=-2,-1,..,2$ is the $\mu$ component of the quadrupole operator. 
The deformation parameters mentioned above are defined as:
\begin{equation}
q_{20}=\frac{\beta_{2}\cos\gamma}{C};\,q_{22}=\frac{\beta_{2}\sin\gamma}{\sqrt{2}C};\,
C=\sqrt{\frac{5}{4\pi}}\frac{4\pi}{3r_{0}^{2}A^{5/3}}
\label{betagamma}
\end{equation}
being $A$ the mass number and $r_{0}=1.2$ fm.
In the present work we have imposed the parity conservation as a self-consistent symmetry of the intrinsic states: $\hat{\mathcal{P}}|\rangle=|\rangle$, being $\hat{\mathcal{P}}$ the parity operator. Therefore, neither negative parity states nor odd-multipole deformation -such as the octupole- degrees of freedom are explored here. Furthermore, these states are invariant under the so-called simplex-x, $\hat{\mathcal{S}}_{x}|\rangle\equiv\hat{\mathcal{P}} e^{-i\pi\hat{J}_{x}}|\rangle=|\rangle$, and the T-simplex-y, $\hat{\mathcal{S}}^{T}_{y}|\rangle\equiv\hat{\mathcal{P}} e^{-i\pi\hat{J}_{y}}\hat{\mathcal{T}}|\rangle=|\rangle$ symmetries, being $\hat{\mathcal{T}}$ the time-reversal operator. The last condition is chosen to have real coefficients in the HFB transformation, and the simplex-x symmetry is very suitable to perform cranking calculations (Eq.~\ref{vap_equat}). 
The set of operators $\lbrace\hat{\mathcal{P}},\hat{\mathcal{S}}_{x},\hat{\mathcal{S}}^{T}_{y}\rbrace$ are the three generators of a subgroup of the more general point group $D^{T}_{2h}$~\cite{PRC_62_014310_2000,PRC_62_014311_2000}. The latter has an additional generator, e.g., the time-reversal operator. Although the use of self-consistent symmetries constrains the inclusion of correlations within the mean-field approach through the spontaneous symmetry-breaking mechanism, they are imposed to reduce the computational burden. In the present case, we will also exploit these self-consistent symmetries to test the performance of the method since they provide non-trivial checks that help to identify possible inconsistencies.
For instance, the choice of the collective coordinates $(\beta_{2},\gamma)$ divides the possible quadrupole deformations in six sextants, depending on the range of the angle $\gamma$~\cite{RingSchuck}. As a result, the values of $\gamma$ equal to $0^{\circ} (60^{\circ})$, $120^{\circ} (180^{\circ})$ and $240^{\circ} (300^{\circ})$ correspond to prolate (oblate) axial deformation and they are related by the different orientations of the principal axes of inertia with respect to the $z$-axis~\cite{RingSchuck}. If $J_{c}=0$, the intrinsic wave functions do not break the time-reversal symmetry and the energy is independent on the orientation of the axes, being all of the sextants completely equivalent. However, if $J_{c}\neq0$, the energy will depend on the orientation of the principal axes of inertia with respect to the intrinsic rotation axis, in our case, the $x$-axis (see Eq.~\ref{vap_equat}). Therefore, the intrinsic states are only invariant under the subgroup of the $D^{T}_{2h}$ point group mentioned above and the sextants are now symmetric only with respect to the $\gamma=(120^{\circ},300^{\circ})$ direction. 

We check this symmetry by performing PN-VAP calculations in the $(\beta_{2},0^{\circ}\leq\gamma\leq360^{\circ})$ plane for three values of the cranking angular momentum $J_{c}=0$ (time-reversal symmetry conserving), 2 and 4, selecting the nucleus $^{32}$Mg as an example. In Fig.~\ref{Figure2}(a)-(c) we show such potential energy surfaces (PES) -first term in the r.h.s. of Eq.~\ref{vap_equat}. Here, the intrinsic states were expanded in nine major spherical harmonic oscillator shells and the number of points included in the mesh of each PES is 502. 
We notice first the equivalence between all of the sextants in the case where the time-reversal symmetry is preserved ($J_{c}=0$, Fig.~\ref{Figure2}(a)). Such a redundancy is reduced to the half of the plane separated by the $\gamma=(120^{\circ},300^{\circ})$ axis for $J_{c}=2$ and 4 (Figs.~\ref{Figure2}(b)-(c)) as expected. We find the absolute minimum of the $J_{c}=0$ PES in the spherical point as it is presumed from the neutron magic number $N=20$. The energy grows more rapidly along the oblate directions than in the prolate ones. Additionally, a second minimum -around 1 MeV higher- is obtained at axial prolate configurations with $\beta_{2}=0.45$. For larger values than $\beta_{2}\approx0.7$ the energy increases quickly also along the prolate lines. For $J_{c}=2,4$ the PES resemble the $J_{c}=0$ one except for the values along $\gamma=120^{\circ}$, where the energy is not as favored as in $\gamma=0^{\circ}$ and $\gamma=240^{\circ}$. The minima of these surfaces appear at such prolate configurations with $\beta_{2}=0.45$, as in the $J_{c}=0$ case but shifted to higher energy values, around $\approx2$ MeV and $\approx3$ MeV for $J_{c}=2$ and 4 respectively. 
\begin{figure}[tb]
\begin{center}
\includegraphics[width=\columnwidth]{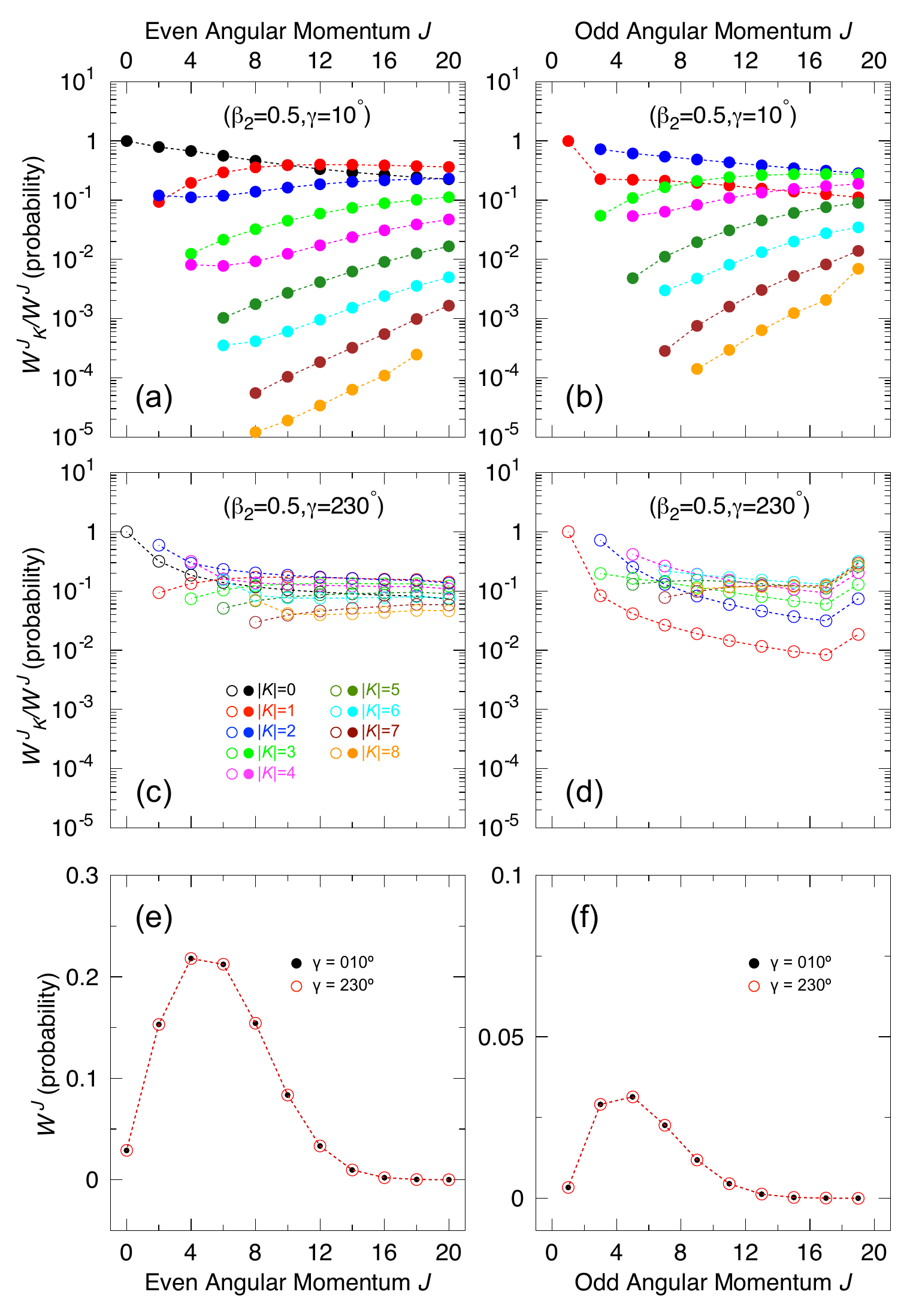}
\end{center}
\caption{(color online) Probability distributions of projections $K$ -$W^{J}_{K}(\beta_{2},\gamma,J_{c})$- for even (left panel) and odd (right panel) values of the angular momentum $J$ for the intrinsic states: (a)-(b) $(\beta_{2},\gamma,J_{c})=(0.5,10^{\circ},4)$ and (c)-(d) $(\beta_{2},\gamma,J_{c})=(0.5,230^{\circ},4)$. Distribution of probabilities of (e) even values and (f) odd values of the angular momentum $J$ -$W^{J}(\beta_{2},\gamma,J_{c})$- for the same intrinsic HFB-type wave functions.}
\label{Figure3}
\end{figure}

The intrinsic many-body states, $|\rangle$, break also the rotational invariance of the hamiltonian and these quantum numbers can be restored by projecting onto good number of particles ($N,Z$) and angular momentum ($J,M$):
\begin{equation}
|JM;NZ;\sigma;\beta_{2},\gamma,J_{c}\rangle=
\sum_{K}g^{J\sigma}_{K(\beta_{2},\gamma,J_{c})}|JMK;NZ;\beta_{2},\gamma,J_{c}\rangle
\label{proj_state}
\end{equation}
where $K=-J,-J+1,...,J$ is the component of the angular momentum in the body-fixed $z$-axis and the states given in the r.h.s. are defined as:
\begin{equation}
|JMK;NZ;\beta_{2},\gamma,J_{c}\rangle=P^{J}_{MK}P^{N}P^{Z}|\beta_{2},\gamma,J_{c}\rangle
\label{PROJ_JKM}
\end{equation}
being $P^{J}_{MK}$ the angular momentum projection operator~\cite{RingSchuck}. Additionally, the coefficients $g^{J\sigma}_{K(\beta_{2},\gamma,J_{c})}$ and the projected energies $E^{J\sigma}_{\beta_{2},\gamma,J_{c}}$ are found variationally by solving the so-called Hill-Wheeler-Griffin (HWG) equations in the $K$-subspace~\cite{RingSchuck}:
\begin{equation}
\sum_{K'}\left(\mathcal{H}_{KK'(\beta_{2},\gamma,J_{c})}^{J;NZ}-E^{J\sigma}_{\beta_{2},\gamma,J_{c}}\mathcal{N}_{KK'(\beta_{2},\gamma,J_{c})}^{J;NZ}\right)g^{J\sigma}_{K(\beta_{2},\gamma,J_{c})}=0
\label{HWG_proj}
\end{equation}
where $\sigma=1,2,...$ labels the possible solutions of such generalized eigenvalue problems and:
\begin{equation}
\mathcal{N}_{KK'(\beta_{2},\gamma,J_{c})}^{J;NZ}=\langle JMK;NZ;\beta_{2},\gamma,J_{c}|JMK';NZ;\beta_{2},\gamma,J_{c}\rangle
\label{norm_kern_proj}
\end{equation}
\begin{equation}
\mathcal{H}_{KK'(\beta_{2},\gamma,J_{c})}^{J;NZ}=\langle JMK;NZ;\beta_{2},\gamma,J_{c}|\hat{H}|JMK';NZ;\beta_{2},\gamma,J_{c}\rangle
\label{hamil_kern_proj}
\end{equation}
are the norm and hamiltonian overlaps. From the above definitions, we can obtain some useful properties of the state 
$P^{N}P^{Z}|\rangle$ such as~\cite{RingSchuck,PRC_59_135_1999}: a) the probability distribution, $W^{J}_{K}(\beta_{2},\gamma,J_{c})$, of finding an eigenstate of the angular momentum  $|JK\rangle$; b) the total probability distribution, $W^{J}(\beta_{2},\gamma,J_{c})$, of finding a value of the angular momentum $J$; and c) the projected energy $E^{J}_{K}(\beta_{2},\gamma,J_{c})$.
\begin{equation}
W^{J}_{K}(\beta_{2},\gamma,J_{c})=\frac{\mathcal{N}_{KK(\beta_{2},\gamma,J_{c})}^{J;NZ}}{\langle\beta_{2},\gamma,J_{c}|P^{N}P^{Z}|\beta_{2},\gamma,J_{c}\rangle}
\end{equation}
\begin{equation}
W^{J}(\beta_{2},\gamma,J_{c})=\sum_{K}W^{J}_{K}(\beta_{2},\gamma,J_{c})
\end{equation}
\begin{equation}
E^{J}_{K}(\beta_{2},\gamma,J_{c})=\frac{\mathcal{H}_{KK(\beta_{2},\gamma,J_{c})}^{J;NZ}}{\mathcal{N}_{KK(\beta_{2},\gamma,J_{c})}^{J;NZ}}
\end{equation}

\begin{figure}[tb]
\begin{center}
\includegraphics[width=\columnwidth]{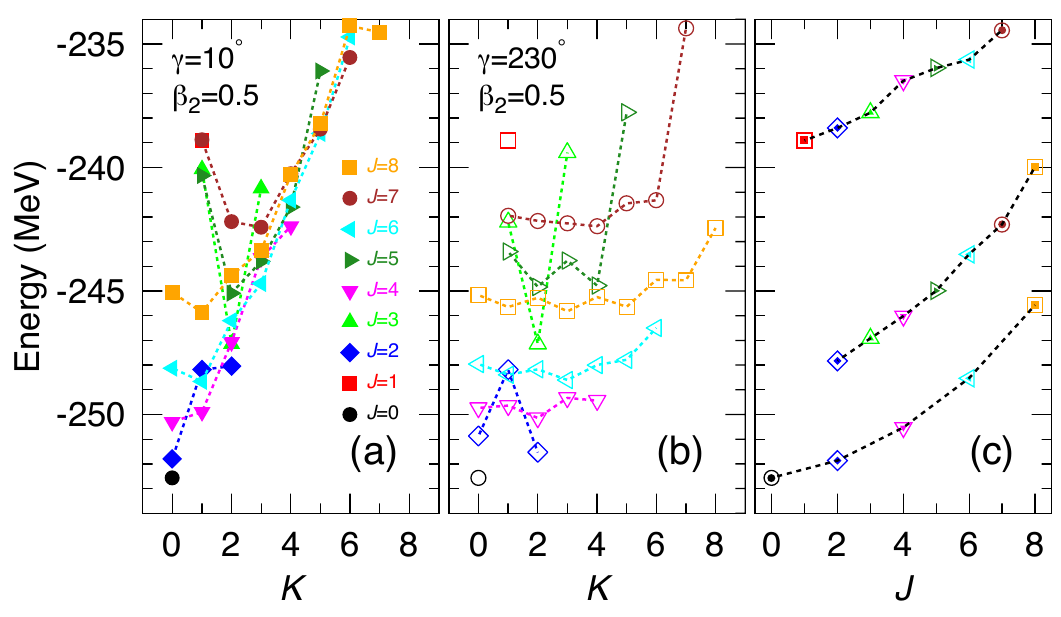}
\end{center}
\caption{(color online) Particle number and angular momentum projected energy overlaps $E^{J}_{K}(\beta_{2},\gamma,J_{c})$ as a function of $K$ for the intrinsic wave functions (a) $(\beta_{2},\gamma,J_{c})=(0.5,10^{\circ},4)$ (filled symbols) and (b) $(\beta_{2},\gamma,J_{c})=(0.5,230^{\circ},4)$ (empty symbols). The last column corresponds to the lowest energies for a given $J$ after $K-$ mixing $E^{J\sigma}_{\beta_{2},\gamma,J_{c}}$ (Eq.~\ref{HWG_proj}) for the same intrinsic states as in (a) and (b).}
  \label{Figure4}
\end{figure}
The decomposition $W^{J}_{K}$ and the energy $E^{J}_{K}$ are quantities that depend on the orientation of the principal axes of inertia with respect to the $(x,y,z)$-axes. Nevertheless, the following properties are deduced from the self-consistent symmetries imposed to the intrinsic states: $W^{J}_{K}=W^{J}_{-K}$, $E^{J}_{K}=E^{J}_{-K}$, and, if $J$ is odd, the $K=0$ component is forbidden.
The dependence on $K$ is removed once the $K$-mixing is performed since the norm $W^{J}$ and the energy $E^{J\sigma}_{\beta_{2},\gamma,J_{c}}$ are scalar quantities~\cite{RingSchuck,PRC_78_024309_2008,PRC_81_064323_2010}. In addition, if $J_{c}=0$ the same probability distribution $W^{J}$ and the same angular momentum projected energy are found in the six sextants of the $(\beta_{2},\gamma)$ plane. However, if the cranking term is non-zero, the $(\beta_{2},\gamma)$ plane is split in two equivalent parts divided by the $(\gamma=120^{\circ},300^{\circ})$ line. 

We now exploit these symmetries to perform consistency tests of the results and check the implementation of the method. Therefore, we select first two intrinsic states, $|\beta_{2}=0.5,\gamma=10^{\circ},J_{c}=4\rangle$ and $|\beta_{2}=0.5,\gamma=230^{\circ},J_{c}=4\rangle$ that are symmetric with respect to the $(\gamma=120^{\circ},300^{\circ})$ line (see yellow squares in Fig.~\ref{Figure2}(c)). We represent the decomposition of those states in components of the angular momentum $J$ and intrinsic $z$-projection $K$ in Fig.~\ref{Figure3}(a)-(d), normalized to the total probability in a given $J$.  Here we observe that the decomposition in $K$ is different depending on the value of $\gamma$. For $\gamma=10^{\circ}$ the probability decreases in general rapidly with increasing $K$ for a fixed value of $J$. Furthermore, the relative weight of the components with large $K$ tends to increase with larger angular momentum $J$, while the $K=0$ component for even $J$ and $K=1,2$ components for odd $J$ slightly decrease. These results are consistent with having the intrinsic long inertia axis nearly along the $z$-axis. On the other hand, the probability for a given $J$ is distributed in a larger number of $K$ components for $\gamma=230^{\circ}$ and these components are much flatter than in the previous case when the angular momentum $J$ is increased. In this case, the intrinsic long inertia axis is almost oriented perpendicular to the $z$-axis.
Nevertheless, the decomposition in $J$ of both states, summing all of the $K$ components, are identical. We observe two separate distributions for even $J$ (Fig.~\ref{Figure3}(e)) and odd $J$ (Fig.~\ref{Figure3}(f)), being the absolute scale larger for the former. The even (odd) distribution probability increases from $J=0$ (1) until the maximum at $J=4$ (5) is reached. Then, $W^{J}$ decreases, obtaining practically zero probability for even (odd) angular momenta larger than 16 (13). 
In  Fig.~\ref{Figure4} ) we represent the $E^{J}_{K}(\beta_{2},\gamma,J_{c})$ energies defined above. We see as in the previous case noticeable differences depending on the $\gamma$ values. For $\gamma=10^{\circ}$ the energies rise rather quickly for large values of $K$ while for $\gamma=230^{\circ}$ the energies are flatter. These differences are completely removed when $K$-mixing is performed through solving the HWG equations (Eq.~\ref{HWG_proj}) as it is shown in Fig.~\ref{Figure4}(c). There, three bands can be distinguished, namely, a ground state rotational band with $\Delta J=2$ built on top of $J=0^{+}_{1}$, and two $\Delta J=1$ bands, being $J=2^{+}_{1}$ and $J=1^{+}_{1}$ the corresponding band-heads.

We can test even further the performance of the angular momentum projection by projecting the whole $(\beta_{2},\gamma)$ plane as it is plotted in Fig.~\ref{Figure2}(d)-(f). 
For the $J_{c}=0$ case, the equivalence between the six sextants is preserved when angular momentum projection is performed. 
However, the angular momentum projected PES attained by restoring the rotational symmetry of the $J_{c}\neq0$ states are symmetric only around the axis $(\gamma=120^{\circ},300^{\circ})$ (Fig~\ref{Figure2}(e)-(f)). In Fig~\ref{Figure2}(d)-(f) only the PES for $(J=J_{c},\sigma=1)$ are shown although the same equivalence is obtained for other values of $(J,\sigma)$.
Apart from the symmetries discussed above, the angular momentum projection modifies significantly the surfaces obtained at the PN-VAP approach. In general, the minima found in the PES at the PN-VAP level are now wider and a slightly larger deformation is obtained whenever the angular momentum is restored. As a matter of fact, these beyond mean-field correlations move the ground state from the spherical point to prolate configurations with $\beta_{2}\approx0.5$ (Fig.~\ref{Figure2}(d)), that was formerly a secondary minimum in the PN-VAP calculation (Fig.~\ref{Figure2}(a)). This effect was already obtained with axial calculations~\cite{NPA_709_201_2002} and is a self-consistent way to obtain the deformed ground state for the nucleus $^{32}$Mg, i.e., as belonging to the 'island of inversion'  with an erosion of the $N=20$ magic number.

The last step in the present SCCM many-body method is the configuration mixing:
\begin{equation}
|JM;NZ;\sigma\rangle=\sum_{\beta_{2},\gamma,K,J_{c}}f^{J\sigma}_{\beta_{2},\gamma,K,J_{c}}|JMK;NZ;\beta_{2},\gamma,J_{c}\rangle
\label{GCM_state}
\end{equation}
Again, the coefficients $f^{J\sigma}_{\beta_{2},\gamma,K,J_{c}}$ in Eq.~\ref{GCM_state}, and the final spectrum, $E^{J\sigma}$, are obtained by solving the general HWG equations:
\begin{equation}
\sum_{\lbrace\alpha'\rbrace}\left(\mathcal{H}_{\lbrace\alpha\rbrace;\lbrace\alpha'\rbrace}^{J;NZ}-E^{J\sigma}\mathcal{N}_{\lbrace\alpha\rbrace;\lbrace\alpha'\rbrace}^{J;NZ}\right)f^{J\sigma}_{\lbrace\alpha'\rbrace}=0
\label{HWG_eq}
\end{equation}
where $\lbrace\alpha\rbrace\equiv\lbrace\beta_{2},\gamma,K,J_{c}\rbrace$ now encodes all the constraints and $K$ in a single index. $\mathcal{H}$ and $\mathcal{N}$ are the energy and norm overlaps respectively:
\begin{equation}
\mathcal{N}_{\lbrace\alpha\rbrace;\lbrace\alpha'\rbrace}^{J;NZ}=\langle JMK;NZ;\beta_{2},\gamma,J_{c}|JMK';NZ;\beta'_{2},\gamma',J'_{c}\rangle
\label{norm_kern}
\end{equation}
\begin{equation}
\mathcal{H}_{\lbrace\alpha\rbrace;\lbrace\alpha'\rbrace}^{J;NZ}=\langle JMK;NZ;\beta_{2},\gamma,J_{c}|\hat{H}|JMK';NZ;\beta'_{2},\gamma',J'_{c}\rangle
\label{ham_kern}
\end{equation} 
\begin{figure}[tb]
\begin{center}
\includegraphics[width=\columnwidth]{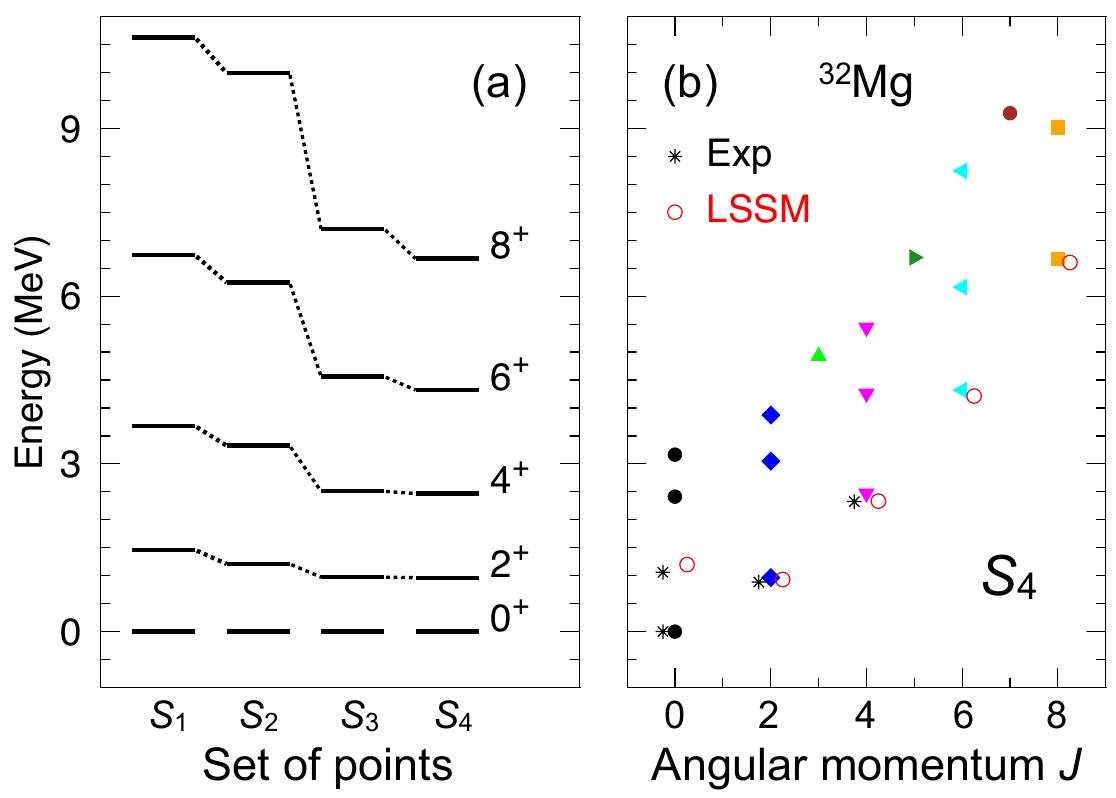}
\end{center}
\caption{(color online) (a) Excitation energies of the yrast states calculated for the nucleus $^{32}$Mg with the GCM method with 17 axial states and $J_{c}=0$ ($S_{1}$), 49 axial+triaxial states and $J_{c}=0$ ($S_{2}$), 81 axial+triaxial states and $J_{c}=0,2$ and 113 axial+triaxial states and $J_{c}=0,2,4$. (b) Excitation energies for the three lowest states for each $J$-value, calculated with the GCM method with 113 axial+triaxial states and $J_{c}=0,2,4$ (full symbols), large scale shell model calculations (open circles) and experimental data (asterisks).}
  \label{Figure7}
\end{figure}
To shed light on the impact of including time-reversal symmetry breaking states in the spectrum, the nucleus $^{32}$Mg has been computed with the GCM method using four sets of intrinsic wave functions. All of them are computed with nine major oscillator shells ($N_{o.s.}=9$) in the working basis~\footnote{We have checked the convergence of the energy spectra by performing axial calculations including up to thirteen major oscillator shells. We have obtained practically the same results as the $N_{o.s.}=13$ ones already with  $N_{o.s.}=7$, thus showing a good convergence of the results with respect to the size of the working basis.}. The simplest one ($S_{1}$) is made of the 17 axial and time-reversal symmetric states. Such states are marked in Fig.~\ref{Figure2}(a) with dots along the $(\gamma=0^{\circ},180^{\circ})$ axis. Then, the $S_{2}$ set is defined by adding 32 more time-reversal conserving states $(J_{c}=0)$ in the $(\beta_{2},\gamma)$ plane (the remaining dots in the same figure). Finally, two more batches of states, $S_{3}$ and $S_{4}$, are established by adding 32 time-reversal symmetry breaking states with $J_{c}=2$, and 32 more with $J_{c}=4$ (see the dots in Fig.~\ref{Figure2}). Therefore, $S_{1}\subset S_{2} \subset S_{3} \subset S_{4}$, being the total number of states in the largest set equal to 113.

The ground state bands calculated with the GCM method implemented with the different sets described above are shown in Fig.~\ref{Figure7}(a). Firstly, the ground state energies obtained for the different calculations are pretty close except for the pure axial case, namely: 

\noindent$E(0^{+}_{1})=-253.056,-253.477,-253.486,-253.498$ MeV for $S_{1}$, $S_{2}$, $S_{3}$ and $S_{4}$ respectively. That shows that the ground state energy is converged with respect to adding time-reversal symmetry breaking components. However, the excited states are more affected by the inclusion of triaxial and cranking states. Hence, we first observe a moderate compression of the spectrum from the axial $(K=0)$ to triaxial calculations with $J_{c}=0$. The decrease in energy is larger with increasing the angular momentum, mainly due to the possibility of having more $K$-mixing in the GCM states. However, the variational space for the excited states are much better explored if time-reversal symmetry breaking is allowed. Therefore, a significant compression of the spectrum is obtained for the $S_{3}$ and $S_{4}$ sets and the differences, once again, are bigger for larger angular momentum. In addition, we can infer that the excitation energies for the $2^{+}_{1}$ and $4^{+}_{1}$ states are already converged with the $S_{3}$ calculation since they do not vary significantly from including $J_{c}=4$ states to the $J_{c}=0,2$ ones. This is not the case for larger values of the angular momentum, where probably intrinsic states with $J_{c}=6,8,...$ should be also included in the GCM. 
For the sake of completeness, the full spectrum computed with the $S_{4}$ set is represented in Fig.~\ref{Figure7}(b). Here, the first two bands display a rotational character, with a parabolic trend in the excitation energies, $0^{+}$ band-heads and $\Delta J=2$ spacing. A third band starting at $2^{+}_{3}$ with $\Delta J=1$ is also obtained, showing a slight odd-even $J$ staggering. In addition, large scale shell model (LSSM) results~\cite{NPA_693_374_2001,PRC_90_014302_2014} and experimental data~\cite{NPA_246_37_1984,PLB_346_9_1995,AIP_Conf_Proc_495_171_1999,PRC_79_054319_2009,PRL_105_252501_2010} are also represented in Fig.~\ref{Figure7}(b). Thanks to the compression of the spectrum produced by the addition of cranking states, a remarkable agreement between the experimental and theoretical values for the $2^{+}_{1}$ and $4^{+}_{1}$ energies is obtained. In addition, the present SCCM calculations predict very similar excitation energies to the LSSM values for the g.s. band. However, the low excitation energy of the $0^{+}_{2}$ state~\cite{PRL_105_252501_2010} is not reproduced here. LSSM calculations have shown that this state is very sensitive to a subtle mixing of spherical 0p-0h and superdeformed 4p-4h configurations~\cite{PRC_90_014302_2014}. In the present framework, the inclusion of pairing fluctuations~\cite{PLB_704_520_2011} and/or explicit quasiparticle excitations could help to solve this problem since the excited $0^{+}$ states are mainly affected by such a degree of freedom, lowering the excitation energies of those states.  
\begin{figure}[tb]
\begin{center}
\includegraphics[width=\columnwidth]{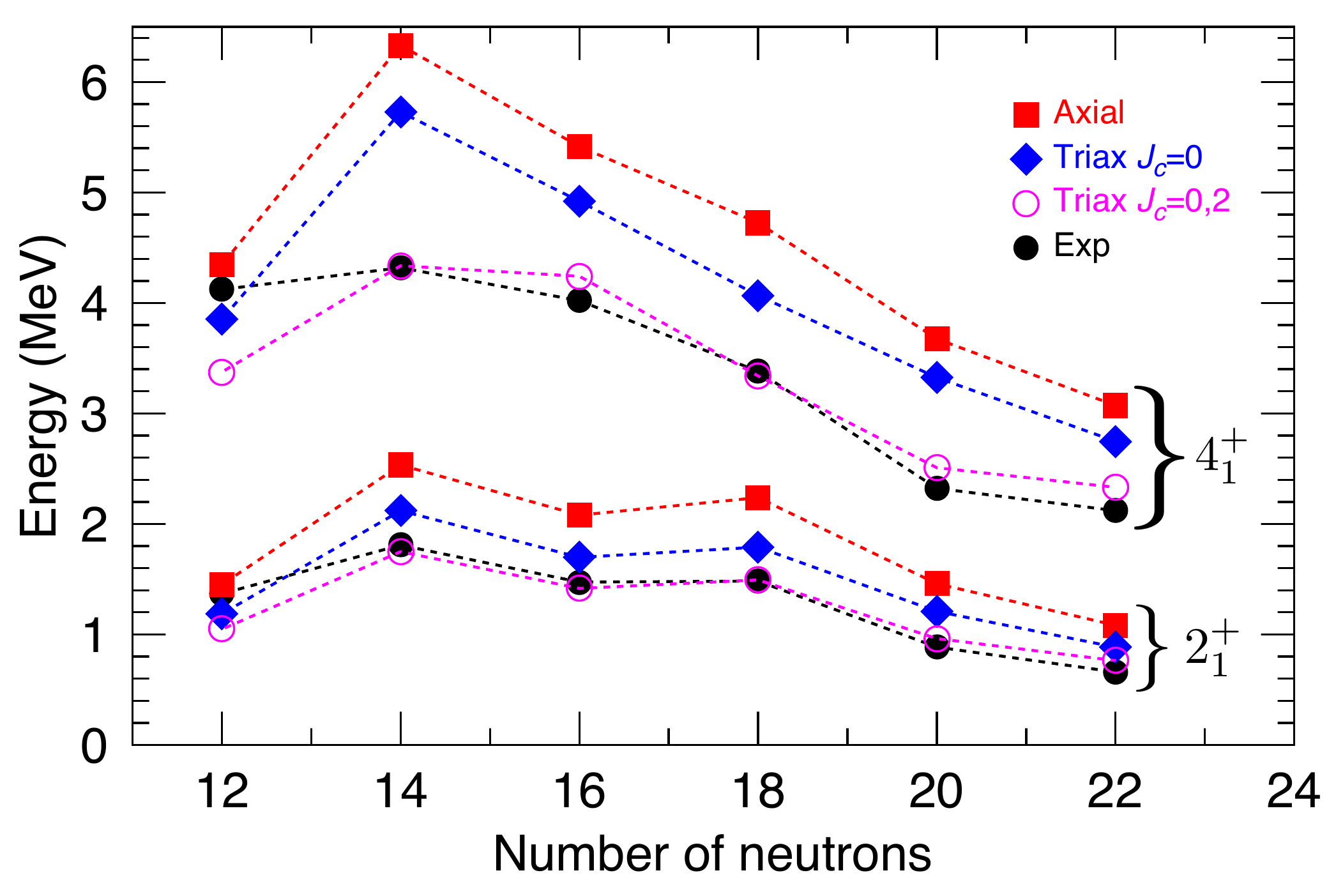}
\end{center}
\caption{(color online) $2^{+}_{1}$ and $4^{+}_{1}$ excitation energies for the Mg isotopic chain calculated with the GCM method including axial states (red squares), axial+triaxial with $J_{c}=0$ states (blue diamonds) and axial+triaxial with $J_{c}=0,2$ states (magenta open dots). Experimental values (black dots) are taken from Ref.~\cite{DataBase} and references therein.}
\label{Figure9}
\end{figure}

Finally, we explore systematically the effect of the inclusion of time-reversal symmetry breaking states in the magnesium isotopic chain $^{24-34}$Mg. The results are obtained with $N_{s.o.}=7$ -the minimum that guarantees a good convergence with respect to the size of the basis in this isotopic chain- and the sets of wave functions defined above $S_{1},S_{2},S_{3}$, i.e., axial and triaxial shapes with $J_{c}=0$ and 2 are included. In Fig.~\ref{Figure9} we plot the excitation energies for the $2^{+}_{1}$ and $4^{+}_{1}$ calculated with these different approaches compared to the experimental values. We see that the axial calculations describe the trends of the experimental data but the energies are largely overestimated. Including the triaxial degree of freedom without breaking the time-reversal symmetry reduces the excitation energies but the predicted values are still too high with respect to the experiments. Finally, adding $J_{c}=2$ states to the GCM set of wave functions compresses further the spectrum and an outstanding agreement with the experimental values is found. The only nucleus where the theoretical values tend to be lower than the experimental ones is the nucleus $^{24}$Mg. Since this is a $N=Z$ nucleus, some alpha clustering and/or proton-neutron pairing correlations could be missing within the present framework which assumes a structure of the intrinsic states given by a direct product of protons and neutrons wave functions. 
However, mixing protons and neutrons to take into account such proton-neutron pairing correlations is beyond the scope of the present study.

In any case, we have to underline that these results constitute the first explicit evidence of the compression of the spectrum when time-reversal symmetry breaking is taken into account in GCM calculations with particle number and angular momentum projection. Global calculations performed with these methods assuming axial symmetry have displayed a systematic overestimation of the $2_{1}^{+}$ excitation energies around a factor $\sim 1.2-1.4$ with respect to the experimental values, both for Skyrme~\cite{PRC_75_044305_2007} and Gogny~\cite{PRC_2015} functionals. The present results show that such a disagreement could be corrected by including triaxial and $J_{c}\neq0$ states in the GCM framework. In fact, the incorporation of $J_{c}$ in the GCM ansatz (Eq.~\ref{GCM_state}) is a generalization of the double projection method of Peierls and Thouless~\cite{NP_38_154_1962,PRC_27_453_1983}. The double projection method is known to provide the exact translational mass in the case of translations by taking as coordinates the position and the linear velocities in a generator coordinate method. We expect, therefore, that the moment of inertia of our theory will be similar to the one provided by the angular momentum projection before variation approach, instead of the Yoccoz moment of inertia given by the angular momentum projection after variation method used in earlier approaches. This expectation is confirmed by our results that provide moments of inertia very close to the experimental ones.

In summary, we have presented the first GCM calculations with particle number and angular momentum projection of HFB-like states considering different quadrupole deformations (axial and triaxial) and intrinsic cranking angular momentum. The performance of the method has been checked by taking advantage of the self-consistent symmetries imposed to the intrinsic many-body states. Since such wave functions were chosen to be eigenstates of a $D^{T}_{2h}$ sub-group generated by the parity, simplex-x and T-simplex-y symmetry operators ($\lbrace\hat{\mathcal{P}},\hat{\mathcal{S}}_{x},\hat{\mathcal{S}}^{T}_{y}\rbrace$), the potential energy surfaces (particle number and particle number plus angular momentum projected) must be symmetric in the $(\beta_{2},\gamma)$ plane with respect to the $(\gamma=120^{\circ},300^{\circ})$ axis. We have checked such a non-trivial property both in individual states and in the whole $(\beta_{2},\gamma)$ plane, taking the nucleus $^{32}$Mg as an example. 
The effect of including incrementally intrinsic states with more symmetries broken in the GCM framework has been also analyzed in $^{32}$Mg and in the magnesium isotopic chain $^{24-34}$Mg. The results have shown that adding ($J_{c}\neq0$) time-reversal symmetry breaking states squeezes notably the spectra due to a better description of the excited states from a variational point of view. Such a compression puts the theoretical values on top of the experimental ones for the lowest $2^{+}$ and $4^{+}$ states in the chain. The next step will be the calculation of electromagnetic properties within the present SCCM approach and some work is in progress along these lines.

\section*{Acknowledgements}
We acknowledge the support from GSI-Darmstadt and CSC-Loewe-Frankfurt computing facilities. T. R. R. thanks A. Poves and F. Nowacki for fruitful discussions and for providing us with the shell model results of $^{32}$Mg.
This work was supported by the Ministerio de Econom\'ia y Competitividad under contracts FPA2011-29854-C04-04, BES-2012-059405 and Programa Ram\'on y Cajal 2012 number 11420 .



\begin{thebibliography}{9}
\bibitem{RMP_77_427_2005}
E. Caurier, G. Mart\'inez-Pinedo, F. Nowacki, A. Poves, and A. P. Zuker, Rev. Mod. Phys. 77, 427 (2005).
\bibitem{PPNP_47_319_2001}
T. Otsuka, M. Honma, T. Mizusaki, N. Shimizu, and Y. Utsuno, Prog. Part. Nucl. Phys. 47, 319 (2001).
\bibitem{RMP_75_121_2003}
M. Bender, P.-H. Heenen, and P.-G. Reinhard, Rev. Mod. Phys. 75, 121 (2003).
\bibitem{Berger84} J.F.~Berger, M. Girod and D. Gogny, 
Nucl. Phys. A 428, 23 (1984).
\bibitem{NPA_671_145_2000}
A. Valor, P.-H. Heenen, and P. Bonche, Nucl. Phys. A 671, 145 (2000).
\bibitem{NPA_709_201_2002}
R. Rodr\'iguez-Guzm\'an, J. L. Egido, and L. M. Robledo, Nucl. Phys. A 709, 201 (2002).
\bibitem{PRL_99_062501_2007} 
T. R. Rodr\'iguez and J. L. Egido, Phys. Rev. Lett. 99, 062501 (2007).
\bibitem{PRC_74_064309_2006}
T. Niksic, D. Vretenar, and P. Ring, Phys. Rev. C 74, 064309 (2006).
\bibitem{PLB_704_520_2011}
N. L. Vaquero, T. R. Rodr\'iguez, J. L. Egido, Phys. Lett. B 704, 520 (2011).
\bibitem{PRC_78_024309_2008}
M. Bender, and P.-H. Heenen, Phys. Rev. C 78, 024309 (2008).
\bibitem{PRC_81_064323_2010}
T. R. Rodr\'iguez, and J. L. Egido, Phys. Rev. C 81, 064323 (2010).
\bibitem{PRC_81_044311_2010}
J. M. Yao, J. Meng, P. Ring, and D. Vretenar, Phys. Rev. C 81 044311 (2010).
\bibitem{PRL_113_162501_2014}
B. Bally, B. Avez, M. Bender, and P.-H. Heenen, Phys. Rev. Lett. 113, 162501 (2014).
\bibitem{NPA_385_14_1982}
K. Hara, A. Hayashi, and P. Ring, Nucl. Phys. A 385, 14 (1982).
\bibitem{NPA_435_477_1985}
E. W\"ust, A. Ansari, and U. Mosel, Nucl. Phys. A 435, 477 (1985). 
\bibitem{PRC_59_135_1999}
K. Enami, K. Tanabe, and N. Yoshinaga, Phys. Rev. C 59, 135 (1999).
\bibitem{PRC_29_1056_1984}
D. Baye and P.-H. Heenen, Phys. Rev. C 29, 1056 (1984).
\bibitem{PRC_76_044304_2007}
H. Zdu\'nczuk, W. Satu\l a, J. Dobaczewski, and M. Kosmulski, Phys. Rev. C 76, 044304 (2007).
\bibitem{RingSchuck}
P. Ring, and P. Schuck, \textit{The nuclear many body problem}, Springer-Verlag,
Berlin, 1980.
\bibitem{PRC_79_044318_2009}
D. Lacroix, T. Duguet, and M. Bender, Phys. Rev. C 79, 044318 (2009).
\bibitem{NPA_696_467_2001}
M. Anguiano, J. L. Egido, and L. M. Robledo, Nucl. Phys. A 696, 467 (2001).
\bibitem{PRC_62_014310_2000}
J. Dobaczewski, J. Dudek, S. G. Rohozi\'nski, and T. R. Werner, Phys. Rev. C 62, 014310 (2000).
\bibitem{PRC_62_014311_2000}
S. Frauendorf, Rev. Mod. Phys. 73, 463 (2001).
\bibitem{NPA_693_374_2001}
E. Caurier, F. Nowacki, and A. Poves, Nucl. Phys. A 693, 374 (2001).
\bibitem{PRC_90_014302_2014}
E. Caurier, F. Nowacki, and A. Poves, Phys. Rev. C 90, 014302 (2014).
\bibitem{NPA_246_37_1984}
D. Guillemaud et al., Nucl. Phys. A 246, 37 (1984).
\bibitem{PLB_346_9_1995}
T. Motobayashi et al., Phys. Lett. B 346, 9 (1995).
\bibitem{AIP_Conf_Proc_495_171_1999}
F. Azaiez et al., AIP Conf. Proc. 495, 171 (1999).
\bibitem{PRC_79_054319_2009}
S. Takeuchi et al., Phys. Rev. C 79, 054319 (2009).
\bibitem{PRL_105_252501_2010}
K. Wimmer et al., Phys. Rev. Lett. 105, 252501 (2010).
\bibitem{DataBase} Brookhaven database, http://www.nndc.bnl.gov.
\bibitem{PRC_75_044305_2007}
B. Sabbey, M. Bender, G. F. Bertsch, and P.-H. Heenen, Phys. Rev. C 75, 044305 (2007).
\bibitem{PRC_2015}
T. R. Rodr\'iguez, A. Arzhanov, and G. Mart\'inez-Pinedo, Phys. Rev. C 91, 044315 (2015).
\bibitem{NP_38_154_1962}
R.E. Peierls and D.T. Thouless, Nucl. Phys. 38, 154 (1962).
\bibitem{PRC_27_453_1983}
J. L. Egido, Phys. Rev. C 27, 453(R)(1983)
\end{thebibliography}
\end{document}